\documentstyle[12pt]{article}
\newcommand{\nc}{\newcommand}
\def\CC{\hbox{\it l\hskip -5.5pt C\/}}
\def\frac#1#2{{\textstyle {#1 \over #2}}}

\nc{\beq}{\begin{equation}}
\nc{\eeq}{\end{equation}}
\nc{\beqa}{\begin{eqnarray}}
\nc{\eeqa}{\end{eqnarray}}
\nc{\lsim}{\begin{array}{c}\,\sim\vspace{-21pt}\\< \end{array}}
\nc{\gsim}{\begin{array}{c}\sim\vspace{-21pt}\\> \end{array}}
\def\ZZ{\hbox{\it Z\hskip -4.pt Z}}

\newcommand{\mysection}[1]{\setcounter{equation}{0}\section{#1}}
\def\no{\nonumber}

\def\&{and}

\begin{document}
\begin{titlepage}

\begin{center}
%October, 1996      \hfill       LPT-96-12\\
%\hfill  hep-th/9606188 \\
\vskip .5 in
{\large \bf $2D-$ Fractional Supersymmetry and Conformal Field Theory for
alternative statistics }
\vskip .3 in
{

  {\bf M. Rausch de Traubenberg}$^{a,}$\footnote{rausch@lpt1.u-strasbg.fr}
  {\bf and P. Simon}$^{b,}$\footnote{simon@lpthe.jussieu.fr}
   \vskip 0.3 cm
   {\it a  Laboratoire de Physique Th\'eorique, Universit\'e Louis Pasteur}\\
   {\it 3-5 rue de l'universit\'e, 67084 Strasbourg Cedex, France}\\ }
  \vskip 0.3 cm
   {\it b  Laboratoire de Physique Th\'eorique et Hautes Energies}\\
    {\it Universit\'e Pierre et Marie Curie,Paris VI et Universit\'e Denis
Diderot, Paris VII} \\
    {\it 2 place Jussieu 75251 Paris cedex 05}\\  
\end{center}

\vskip .5 in
\begin{abstract}

Supersymmetry can be consistently  generalized in one and two dimensional
spaces, fractional supersymmetry being one of the possible extension.
$2D$ fractional supersymmetry of arbitrary order $F$ is explicitly constructed
using an adapted superspace formalism. This symmetry   connects
the fractional spin states ($0,{1 \over F}, \cdots,{ F-1 \over F}$). Besides
the  stress momentum tensor, we obtain a conserved  current of spin
($1 + { 1 \over F})$. The coherence of the theory imposes strong constraints
upon the commutation relations of the modes of the fields.
The creation and annihilation operators turn out to generate 
alternative statistics, currently referred as quons in the literature. We 
consider, with a special attention,  the consistence of the algebra, on the 
level of the  Hilbert space and the Green functions. The central charges are 
generally   irrational  numbers
except  for the particular cases $F=2,3,4$. A natural classification 
emerges according to the decomposition of $F$ into its product of prime 
numbers leading to sub-systems with smaller symmetries.

\end{abstract}

\vspace{2cm}
\begin{center}
Accepted for publication in Nucl. Phys. {\bf B.}
\end{center}

\noindent
Pacs: 11.25.Hf; 03.65.Fd; 11.10.Kk; 02.90.+p\\\no
Keywords: Fractional supersymmetry; Conformal invariance; Algebraic field theory;\\\no $q-$deformation; Integrable models.
\end{titlepage}

%%%%%%%%%%%%%%%%%%%%%%%%%%%%%%%%%%%%%%%%%%%%%%%%%%%%%%%%%%%%%%%
\renewcommand{\thepage}{\arabic{page}}
\setcounter{page}{1}
%\mysection{Introduction and Pre-regulation}

 \vskip .5 truecm
 \mysection{Introduction}
 
After the work of Belavin, Polyakov and Zamolodchikov \cite{bpz}, conformal 
invariance has become a powerful tool for the description of $2D$ critical
phenomena. Then, one of the main task would be to have a systematic 
classification of conformal systems. The first attempt in this direction has 
been challenged by Friedan, Qiu and Shenker (FQS) \cite{fqs1} by imposing 
unitarity. Then, it has been proved that, if we enlarge the $2D$ symmetry, 
we can go beyond the discrete series found by FQS. For instance, the 
superconformal extension of the Virasoro algebra leads to other unitary 
series \cite{fqs2}. Bosonic extensions of conformal symmetries are also 
allowed as for instance the Kac-Moody algebra \cite{go} or the $W_n$ 
algebras \cite{w}. The former contains conserved current of spin one whereas 
the latter of spin $n$.\\
However, due to the special feature of $2D$, one is allowed to define fields 
that are neither fermions nor bosons but of fractional conformal weight. 
The parafermions introduced by Fateev and 
Zamolodchikov possess a rational conformal weight \cite{fz}. Those fields 
are the basic building blocks of fractional superconformal Virasoro (FSV) 
algebra   \cite{fv,fsv} and lead naturally to  conserved current of 
fractional conformal weight.
Furthermore, there is an essential difference with the $W_n$ algebras in so 
far as
FSV close through quadratic relations but involve a non linear dependence 
of the fields. Therefore, it is obviously not a Lie or a super-Lie algebra. 
As the fractional Virasoro algebra is concerned, we get quadratic relations 
but rational power are involved in the OPE. Because cuts are involved, the 
theory 
appears to be non-local.\\
All known conformal field theories (CFT)
can be obtained within the framework of the GKO \cite{gko} coset
construction where appropriate Kac-Moody algebras are involved. Let us point 
out that the GKO construction can be applied with all kinds of affine-Lie 
algebras.
Therefore, following this line, one can build other extensions than the ones 
given here above.\\
The starting point  of the present article is rather different; we take 
advantage of the possibility to generalize supersymmetry in one and two 
dimensions. Namely, we use appropriate extensions of Grassmann variables 
\cite{gga1,gga2} to build explicitly an invariant action (by the help of the 
natural extension of the usual superfield). To our knowledge, this is the 
first 
time, that a CFT is obtained with such variables. This symmetry named 
fractional 
supersymmetry (FSUSY)
has been introduced by Durand \cite{d}. It can be seen as  the 
$F^{th}-$root of the time translation in $1D$
\cite{d,fsusy1d,fr} or of the conformal transformation in $2D$ 
\cite{fsusy2d,ssz,prs}. A group theoretical justification of this symmetry was
given in \cite{am,fr}. In a former paper, we have particularized the case 
$F=3$ and stressed on the underlying superspace formalism  \cite{prs}.
The Virasoro algebra is then extended and, besides   the stress-energy 
tensor, we obtain  a conserved current of conformal weight(spin) $(1+ {1 \over 
F})$. Consequently, in addition to the scalar field, we introduce  
primary fields  of conformal weight $ {1 \over F}, \cdots , {F-1 \over 
F}$. It turns out that fractional supersymmetry is the symmetry which connects 
those ${1 \over F}$-integer spin states. As already mentioned in our previous
paper, this extended Virasoro algebra has nothing to do with the fractional
supervirasoro (FSV) one \cite{fv} where a spin $(1+ {1 \over F})$ conserved 
current 
is  already present. The main reason of this difference  is 
that ours closes through local (but non-quadratic) relations whereas FSV closes
 with non-local (but quadratic) ones.\\
\no     
In this article, we are generalizing our previous results
to fractional supersymmetry   of arbitrary order $F$ and take  a special 
emphasis on the consistence of the algebra, especially on the level of the 
Hilbert space and the Operator  Product Expansion (OPE). From this analysis, 
alternative
statistics, currently referred as quons in the literature, emerges in a 
natural way, generalizing fermionic statistics.\\
The paper is organized as follow.
In section $2$, we recall the foundations of the generalized Grassmann algebra 
which allows to build an extension of the Virasoro algebra. Section $3$ is 
devoted to an explicit construction of the FSUSY action, by introducing a 
superfield
of conformal weight $0$. The components of the latter are fields of conformal 
weights $0,\cdots,{F-1\over F}$. The section $4$ contains one of the main 
results of this paper. By developing the various fields in mode expansions, 
we 
find that the coherence of the algebra {\it imposes} special commutation 
relations between the modes. We recover in a natural way the quons algebra, 
previously introduced by Greenberg and Mohapatra \cite{quon}, 
in order to obtain new 
statistics with a small violation of the Pauli principle. The essential 
feature 
of this algebra is the absence of bilinear relations between two creators and 
two annihilators. A consequence of this oscillator structure is an extension 
of 
the Wick theorem. Then, the section $5$ is devoted to the determination of the 
OPE. Some general rules are given in order to ensure  the 
associativity of the algebra at the level of the four-point functions.
Section $6$ shows that  a natural classification 
emerges according to the decomposition of $F$ into its product of prime 
numbers. In other words, the general case can be obtained from the study of 
$F$, when $F$ is a pure prime number.
Then, it becomes straightforward to extend this result to any  $F$.
 Moreover, in addition to the FSUSY symmetry,  we prove for $F=F_1\times F_2$, 
that
this action is also invariant under $F_1$ or $F_2$-supersymmetric 
transformations. Finally, section $7$ contains a summary of our results and 
some
future perspectives for this work.

\mysection{Fractional superconformal algebra}

In this section, 
we need first to recall briefly the underlying algebra which allows to
define FSUSY. The basic fields live in a {\it ad hoc} extension of the complex 
plane, namely
$(z,\theta_L,\bar{z},\theta_R)$ with $\theta_L,\theta_R$ two real generalized  
Grassmann variables \cite{gga1,gga2}. We also introduce the associated 
derivatives $\partial_L, \delta_L, \partial_R, \delta_R$.
They satisfy the basic algebraic relations
\eject
\begin{eqnarray}
\label{eq:alg}
&&\partial_\theta \theta - q \theta \partial_\theta =  1 \nonumber \\
&&\delta_\theta \theta -  q^{-1} \theta \delta_\theta   = 1 \nonumber \\
&&\theta^F=0 \ \ \ \ d_\theta^F=0    \\
&&\partial_\theta \delta_\theta = q^{-1} \delta_\theta \partial_\theta,   
\nonumber
\end{eqnarray}
\noindent
with $\theta = \theta_L$ or $\theta_R$, $d=\partial$ or  $\delta$ and  
$q$  is a primitive {\it F-th} root of unity. Without lost of generality, $q$
can be chosen as $q=\exp{({2i \pi \over F})}$. A consequence of relations 
(\ref{eq:alg}) is an adapted Leibniz rule leading to $\partial_\theta \left(
\theta^a \right) = \{a\} \theta^{a-1}$ with $\{a\} = {1-q^a \over 1-q}$ 
(with the derivative $\delta_\theta$ we would have obtained the same result 
with
the substitution $q \to q^{-1}$).
The relations which mix the $L-$ or $R-$ movers  are

\begin{eqnarray}
\label{eq:algb} 
 \theta_L \theta_R = q\theta_R \theta_L, ~~~
&&d_L d_R = q d_R d_L\\ 
d_L \theta_R = q^{-1}\theta_R d_L,
&&d_R \theta_L = q \theta_L d_R. \nonumber 
\end{eqnarray}

\noindent
From this  algebra,  we can build the generators ($Q$) and the covariant
derivatives ($D$) of   FSUSY 

\begin{eqnarray} 
\label{eq:fsusy}
Q_L&=& \partial_L + {(1-q) \over F}^{F-1} \theta_L^{F-1} \partial_z 
\nonumber \\
D_L&=& \delta_L + {(1-q^{-1})\over F}^{F-1}  \theta_L^{F-1} \partial_z 
\nonumber \\
Q_R&=& \delta_L + {(1-q^{-1})\over F}^{F-1}  \theta_R^{F-1} \partial_{\bar z} 
\\ 
D_R&=& \partial_L + {(1-q)\over F}^{F-1}  \theta_R^{F-1} \partial_{\bar z},
\nonumber
\end{eqnarray}
which fulfill
\begin{eqnarray}
\label{eq:fsusyb} 
Q_L^F=D_L^F=\partial_z~~~~~&&Q_R^F=D_R^F=\partial_{\bar z}\nonumber \\
Q_L D_L=q^{-1}D_L Q_L &&Q_R D_R=q D_R Q_R.
\end{eqnarray}
\noindent
We want to stress that the generalized Grassmann variables are real, therefore,
in the superspace $(z,\theta_L,\bar{z},\theta_R)$, the two generalized 
Grassmann variables are {\it not} complex conjugate from each other. We then 
get an heterotic extension of the complex plane. However, the underlying 
algebra is stable under the composition of the complex conjugation and the 
permutation of $R$ and $L$. This explains why $Q_L,D_L$ and $Q_R,D_R$ have a 
different treatment (see eq. (\ref{eq:fsusy})). We refer to ref. \cite{prs} 
for more details.
All these relations (\ref{eq:fsusyb}) can be obtained directly using
(\ref{eq:alg}--\ref{eq:algb}) or more easily from the faithful matrix 
representation given in \cite{hq}.
The generators defined in (\ref{eq:fsusyb}) can be extended, following 
Durand \cite{d2}, to
\beqa
\label{eq:vir}
L_n&=&z^{1-n} \partial_z -{1 \over F}(n-1) z^{-n} {\cal N},~~~~
n \in \ZZ \nonumber \\
G_r&=&z^{{1 \over F} -r}(\partial_{\theta} +{(1-q) \over F}^{F-1}
 \theta^{F-1} \partial_z) - \\
&&{(1-q) \over F}^{F-1}(r-{1 \over F})z^{{1 \over F}
-r-1}\theta^{F-1} {\cal N},~~~~ r \in \ZZ +{1 \over F}\nonumber,
\eeqa
\noindent
with $ {\cal N}$ the number operator which equals to
 $ {\cal N}=\sum\limits_{i=1}^{F-1}{ (1-q)^i \over (1-q^i)} \theta^i
\partial_\theta^i$. The $L$ and the $G$ generate the
 fractional-super-Virasoro algebra, without central extension,  
\beqa
\label{eq:viral}
\left[L_n,L_m\right]& =& (n-m)L_{m+n} \nonumber \\
\left[L_n,G_r\right] &=& ({n \over F}-r)G_{n+r} \\
\left\{G_{r_1},\cdots,G_{r_F}\right\}&=& L_{r_1 + \cdots + r_F}, \nonumber
\eeqa 
\noindent
where $\left\{G_{r_1},\cdots,G_{r_F}\right\}$ is defined by ${1 \over F!}(
\mathrm{~sum~ over~ all~ the~ permutations~ of~ }G_{r_1},\cdots,G_{r_F} )$.
It then appears that FSUSY is a natural generalization of supersymmetry  
because  it corresponds to the $F-th$ root of the Virasoro generators. 
Of course, the  antiholomorphic part is built in the same way.

\mysection{Construction of the action}

We take, in this section, $F$ as a prime number. We will see later that it 
constitutes a generic case.  First of all, 
we recall 
briefly the basic points which lead from the algebra 
(\ref{eq:fsusy}--\ref{eq:fsusyb}) to 
an invariant FSUSY action. It is interesting to notice that most  
results, in usual  supersymmetric theories \cite{susy} can be extended to 
FSUSY. We are able to build an invariant action in FSUSY, extending
the usual superspace formulation involved in supersymmetric theories  (by the 
help of the generalized Grassmann variables (\ref{eq:alg}--\ref{eq:algb}) ).
 
Therefore, a basic superfield decomposes (in the fractional superspace 
$(z,\theta_L,\bar{z},\theta_R)$) as 

\begin{equation}
\label{eq:field}
\Phi(z,\theta_L,\bar z, \theta_R) \sim \sum \limits_{a,b=0}^{F-1} 
\theta_L^a  \theta_R^b \psi_{{a \over F},{b \over F}}(z,{\bar z}).
\end{equation} 

\noindent
In this multiplet, we have three kinds of fields : the  holomorphic ones  
$\psi_{{a \over F},0}$, the  antiholomorphic ones   $\psi_{0,{a \over F}}$, 
and the auxiliary
fields $\psi_{{a \over F},{b \over F}}$ with $a$ and $b \neq 0$. 
The various components of $\Phi$ generalize the concept of boson/fermion 
and have non-trivial $\ZZ_F-$graduation. The $\theta$ field is then a 
``graduation counter'' and we have the $q-$muta\-tions relations \cite{d,am} 
\begin{eqnarray}
\label{eq:qmut}
(\psi_{{a \over F},{b \over F}})^F &=&0 \nonumber \\
\theta_L \psi_{{a \over F},{b \over F}} &=&q^{-(a+b)} 
\psi_{{a \over F},{b \over F}} \theta_L \\ 
\theta_R \psi_{{a \over F},{b \over F}} &=&q^{-(a+b)} 
\psi_{{a \over F},{b \over F}} \theta_R.\nonumber
\end{eqnarray}
\noindent
Finally, from integration rules ($\int d \theta=\left({d\over 
d\theta}\right)^{F-1}$) \cite{gga2}, we can define the FSUSY invariant action
\beq
\label{faction0} 
\int  d \theta_L d \theta_R \left[\overrightarrow{D_L} \Phi\right] 
\left[\Phi \overleftarrow{D_R}\right], 
\eeq
where $D_L$ (respectively $D_R$) acts from the left (resp. the right).
This nice generalization of SUSY comes from the definition of the
superfield $\Phi$ and the property that the covariant derivative, say $D_L$,
commutes with the FSUSY-transformation $\epsilon_L Q_L$ (see below). In the 
sequel, we just consider the holomorphic part of the action.
The antiholomorphic part is totally similar and the auxiliary fields are 
irrelevant for our study. $D$ and $\theta$ stand 
respectively for $D_L$ and $\theta_L$. Then, the remaining part of the action 
can be equivalently written (with adapted normalizations coming from 
integration
and derivation over $\theta$)
\beq
\label{eq:faction}
{\cal A}=  \left( { 1 -q \over 1-q^{-1}} \right)^{F-1}
\int  d \theta  D \Phi \partial_{\bar z} \Phi.
\eeq
\noindent
The action (\ref{faction0}) is invariant under the FSUSY transformations 
for the left and 
right movers (this is the action we consider in superstring theory for $F=2$).
On the other hand, (\ref{eq:faction}) is only FSUSY invariant  for the 
left movers (it is the action we use in heterotic string when $F=2$). In a 
similar manner, we
can construct a theory which is $F_L$ (respectively $F_R$) invariant by 
choosing $\theta_L$ (respectively $\theta_R$) a generalized Grassmann 
variable of order $F_L$ (respectively $F_R$). Such extension involves 
auxiliary fields. A general study including all the auxiliary fields has been 
performed for $F_L=F_R=3$ in
\cite{prs}.

With the holomorphic part of the previous superfield defined in 
(\ref{eq:field})  $\Phi$ can be written 
$$\Phi(z,{\bar z},\theta)=X(z,{\bar z})+\sum \limits_{a=1}^{F-1} 
q^{{a^2\over 2} }\theta^a   \psi_{{a\over F}}(z).$$
Using the $q-$mutation (\ref{eq:qmut}) and the integration  rule upon the 
generalized Grassmann variables \cite{gga2}, the action (\ref{eq:faction}) 
yields to
%\footnote{This action have some minor (sign) differences
%with respect to \cite{prs}, this corresponds to a change of normalization
%in the covariant derivative $D$.}
\beq
\label{eq:action}
{\cal A}=\partial_z X(z) \partial_{\bar z}X(z)+ { F \over (1- q ^{-1})^F}
\sum \limits_{a=1}^{F-1}
 \left(q^{-a}-1 \right)  \psi_{{a \over F}}(z) \partial_{\bar z}
\psi_{{F-a \over F}}(z).
\eeq
\noindent 
This action is the natural generalization of the supersymmetric ones, 
and has already been considered for the $F=3$ case  \cite{fsusy2d,ssz,prs}
or even for arbitrary $F$ in one dimension \cite{d}.

In a way analogous to \cite{prs}, we can define a path integral, and the 
non-vanishing two-point Green functions are then

\beqa
\label{eq:green}
<\psi_{{F-a\over F}}(z) \psi_{{a\over F}(w)}>&=& {(q-1)\over F}^F {1\over 
q^{-a}-1}
~{1\over z-w}  \nonumber\\
<\psi_{{a\over F}}(z) \psi_{{F-a\over F}}(w)>&=& {(q-1)\over F}^F {1\over 
q^{a}-1}
~{1\over z-w} \\ 
<X(z) X(w)> &=& -\ln(z-w). \nonumber
\eeqa
These equations explicitly show that $\psi_{{a\over F}}(z)$ is the conjugate of
$\psi_{{F-a\over F}}(z)$. This comes from the peculiar structure of the 
Lagrangian, where these two fields have to be coupled in order to have an 
action of conformal weight $0$. When $F=2$, the situation appears rather 
different because fermionic fields are self-conjugate. 

\mysection{Oscillators}
\subsection{Oscillator algebra} 
The solutions of the equations of motion allow to develop the various fields
in terms of the Laurent expansions: $\psi_{{a \over F}}(z) = \sum_r \psi_{a,r}
z^{-r - {a \over F}}$, the index $r$ belonging to $\ZZ + b/F$, $b$ depending of
the boundary conditions of the fields (see \cite{prs} for more details).

\noindent
In this context, nothing can be said 
{\it a priori} on the $q-$mutation relations of the fields. 

\noindent
 For the sake of simplicity, we
consider only the case $F=3$. We thus have 
\beqa
\psi_{{1 \over 3}}(z)&=&\sum_r \psi_{1,r} z^{-r - {1 \over 3}} \\
\psi_{{2 \over 3}}(z)&=&\sum_s \psi_{2,s} z^{-s - {2 \over 3}}.\nonumber
\eeqa
The notations for the indices of the field are different from our previous 
paper
\cite{prs}.
The properties of the underlying algebra induce
strong constraints upon the various modes of the fields and allow to
define unambiguously the vacuum.
The first constraint comes from the possibility to obtain the two-point
Green function, using the mode expansion of the fields. We thus set

\beq
\psi_{1,r} |0> = 0~;~
\psi_{2,s} |0> = 0, {\mathrm with~} r,s > 0. 
\eeq  

\noindent
We then make the following identification
\beqa
\psi_{1,r} \equiv a_r&&\psi_{2,s} \equiv b_s,~~r,s>0\\
\psi_{1,-s} \equiv b^{\dag}_s&&\psi_{2,-r} \equiv a^{\dag}_r,~~r,s>0~.\nonumber
\eeqa
Therefore, the fields can be written as
\beqa
\label{eq:champs}
\psi_{{1 \over 3}}(z) &=& \sum \limits_{s >0} \left\{ b_s^{\dag} z^{s - 1/3}+
a_s z^{-s-1/3} \right\} \equiv \psi_{ 1/3 <}(z) +  \psi_{ 1/3 >}(z) \\
\psi_{{2 \over 3}}(z) &=& \sum \limits_{r >0} \left\{ a_r^{\dag} z^{r - 2/3}+
b_r z^{-r-2/3} \right\} \equiv \psi_{ 2/3 <}(z) +  \psi_{ 2/3 >}(z). \nonumber
\eeqa
\noindent
 Then from (\ref{eq:green}) with $F=3$, we automatically
get 

\beqa
\label{eq:osc}
<\psi_{1 \over 3}(z) \psi_{2 \over 3}(w)> &=& 
<\psi_{{1 \over 3}>}(z) \psi_{{2 \over 3}<}(w)>=
\sum_s<a_s a^{\dag}_s>
\left( {w \over z}\right)^s z^{-2/3}w^{-1/3} \sim
{-q \over z-w} \\
<\psi_{2 \over 3}(z) \psi_{1 \over 3}(w)>& =&
 <\psi_{{2 \over 3}>}(z) \psi_{{1 \over 3}<}(w)>=
 \sum_r<b_r b^{\dag}_r> 
\left( {w \over z}\right)^r z^{-1/3}w^{-2/3} \sim
{q^2 \over z-w },\nonumber
\eeqa
\noindent

if we assume
\beqa
\label{eq:qab1}
a_r a^{\dag}_s - q a^{\dag}_s a_r& =& -q\delta_{rs} \\
b_r b^{\dag}_s - q^2 b^{\dag}_s b_r& = &q^2 \delta_{rs}.  \nonumber
\eeqa
\noindent
Notice that the operators $a,a^{\dag}$ and $b,b^{\dag}$ are not Hermitian 
conjugate from each-other as could have been suggested by our notation (see 
(\ref{eq:qab1})).
Similarly, from the other two-point functions which vanish
$<\psi_{{1 \over 3}}\psi_{{1 \over 3}}>$ and 
$<\psi_{{2 \over 3}}\psi_{{2 \over 3}}>$, we get
  the $q-$muta\-tion relations
  
\beqa
\label{eq:qab2}
a_rb_s^{\dag}&-&q^2 b_s^{\dag}a_r=0\\
b_ra_s^{\dag}&-&q a_s^{\dag}b_r=0. \nonumber
\eeqa
Strictly speaking, at this point
the power of $q$, 
inside the $q-$mutation relations is not fixed,  however, it can be fixed by 
the OPE (to have the right OPE of the stress-energy tensor with the fields). 
Furthermore, these relations are not surprising because 
the two
fields $\psi_{{1 \over 3}},\psi_{{2 \over 3}}$ are conjugated from each other
(see the two-point functions).
Now, it remains {\it a priori} to fix the relations between two annihilators 
or two creators.
If we assume that we have bilinear relations between two $a,b$ or
$a^{\dag}, b^{\dag}$, we get an incoherence in the Hilbert space
(for instance  by calculating  $a_r a^{\dag}_r b^{\dag}_s |0>$ for $r \ne s$
in two different ways). Hence, we are obliged to let these relations 
unconstrained.
Therefore, the natural structure emerging from a quantization of $2D$ FSUSY 
appears to be the quon algebra introduced and developed by Greenberg and
Mohapatra \cite{quon}.
In fact, it should be noticed that in our case, we have $|q|=1$, whereas 
for the quons, $q\in [0,1]$. However, we will see in the next subsection that 
we 
are able to build a positive and definite Hilbert space with $a^{\dag}_r$ and 
$b^{\dag}_r$. To be complete, we have to specify that
\beq
\label{qab3}
(a_r)^3=(a_r^{\dag})^3=(b_r)^3=(b_r^{\dag})^3=0
\eeq
The equations (\ref{eq:qab1},\ref{eq:qab2},\ref{qab3}) are the quantum 
versions of 
$$\left(\psi_{{1 \over 3}}\right)^3=\left(\psi_{{2 \over 3}}\right)^3=0$$

We now define the normal ordering prescription as usual
\beq
\label{eq:no}
:\phi_{1 }(z) \phi_{2 }(z): \equiv
\lim \limits_{z \to w}\left[\phi_{1 }(z) \phi_{2 }(w)-
<\phi_{1}(z) \phi_{2 }(w)>\right] ,
\eeq
\noindent
with $\phi_i$  an arbitrary field.
Consequently, we obtain in a straightforward way
\beqa
\label{qchamp1}
:\psi_{{1 \over 3}<}(z)\psi_{{2 \over 3}>}(z): &=&
q :\psi_{{2 \over 3}>}(z)\psi_{{1 \over 3}<}(z):\nonumber\\
:\psi_{{2 \over 3}<}(z)\psi_{{1 \over 3}>}(z): &=&
q^2 :\psi_{{1 \over 3}>}(z)\psi_{{2 \over 3}<}(z):\\
\label{qchamp2}
:\psi_{{2 \over 3}<}(z)\psi_{{2 \over 3}>}(z): &=&
q :\psi_{{2 \over 3}>}(z)\psi_{{2 \over 3}<}(z):\nonumber\\
:\psi_{{1 \over 3}<}(z)\psi_{{1 \over 3}>}(z): &=&
q^2 :\psi_{{1 \over 3}>}(z)\psi_{{1 \over 3}<}(z):
\eeqa
\noindent
Some remarks are now in order here.\\
Firstly, due to  the quon algebra, we have a weaken Wick theorem,
where only relations between positive and negative frequencies are known.
Let us notice that this peculiar structure will not affect the calculations of 
the  correlations functions as we will see further.
\\Secondly, when we perform for example $:\psi_{{1 \over 3}}^2(z):\psi_{{2 \over 
3}}(w)$, we first return to the definition of our Wick theorem (\ref{eq:no}),
namely $:\psi_{{1 \over 3}}^2(z):=\lim\limits_{\epsilon\to 0}\psi_{{1 \over 
3}}(z+\epsilon)\psi_{{1 \over 3}}(z)$,
then do all the possible contractions and finally take the limit $\epsilon\to 
0$, as it should be.

This is strongly different from the case $D=1$ where the third power of the
$\psi$ fields vanishes, even after quantization \cite{fr} imposing
$\partial_t \psi_1(t) \psi_1(t) = q^{\pm} \psi_1(t)\partial_t \psi_1(t)$. 
When $D=2$,
the third power is not zero so $\partial_z\left(:\psi_{{1 \over 3}}(z)^3:
\right)  \ne 0$
so the fields $\psi_{{1 \over 3}} $ and $\partial_z\psi_{{1 \over 3}}$ cannot
$q-mute$. In the case $F=2$, this subtlety never appears, because after 
quantization,  a Grassmann variable becomes a Clifford one, hence the square
of a fermion is one. 

For arbitrary $F$, all these results can be extended as follow: we have 
${F-1\over 2}$ sectors $(\psi_{{a\over F}},$ $\psi_{{F-a\over F}}) $ with the 
substitution $q\longrightarrow \exp{{2i\pi a\over F}}$.  
Note that the fields $\psi_{{b \over F}}$ and $\psi_{{a \over F}}$ (with 
$b \neq (F-a)$)   commute without the normal ordering prescription 
($: ~~~:$) because they come from different graded sectors.\\
Moreover, before ending this subsection, it is worth noticing that if we 
modify the $q-$muta\-tions relations such that
\beq
a_ra_s^{\dag}-qa_s^{\dag}a_r=k_r(\Delta)\delta_{rs}
\eeq
we obtain
the two-point function
\beq
<0|\psi_{{1 \over 3}}(z)\psi_{{2 \over 3}}(w)|0>={1\over z^{\Delta}}\sum
\limits_{n\ge 0}k_n(\Delta)({w\over z})^n\sim{1\over (z-w)^{\Delta}}.
\eeq
This leaves open the connection between our approach and the parafermions, 
where correlations functions involve fractional power. Naturally, the 
Lagrangian and the superspace have to be modified subsequently.

\subsection{Hilbert space and quons}

As we have mentioned previously, the operators $a, a^{\dag}$ and $b, b^{\dag}$
are not conjugate from each other, however it is possible to make a 
redefinition
such that they become conjugate. Before doing such a transformation, we would 
like to give
some basic properties of the algebra (\ref{eq:qab1},\ref{eq:qab2}).
The peculiarity of such type of algebras
is that $a_r^{\dag} a_s^{\dag} |0> \neq a_s^{\dag} a_r^{\dag} |0>$ for 
$r \neq s$ etc. Consequently, such states decompose into irreducible 
representations of the permutation (eventually the braided) group as it is the
case for the parafermions introduced by Green \cite{para}~\footnote{These
parafermions are different from the ones introduced by Fateev and 
Zamolodchikov.}.
So we have to be careful with the position of the various operators
of creation. We note 
$\left(a_{r_1}^{\dag}\right)^{k_1} \cdots \left(a_{r_i}^{\dag}\right)^{k_i} |0>
\sim |(k_1)_{r_1};\cdots,(k_i)_{r_i}>$. Let  us  now recall and give some 
properties of the representation of the underlying algebra

\begin{enumerate}
\item No bilinear relations can be consistently postulated among two
creators or two annihilators.
\item Any operator of creation cannot act more than $F-$times on the vacuum.
It means that if we consider the state

$$ |h>= A_0 \left(a_r^{\dag}\right)^{k_1} A_1 \cdots A_{i-1} 
\left(a_r^{\dag}\right)^{k_i} A_{i} |0>,$$

\noindent
with $A_0, \cdots, A_i$  arbitrary products of creators different from 
$a_r^{\dag}$, we have $a_r^{\dag} |h> =0$ if $k_1 + \cdots +k_i=F-1$.
In other words, the representation decomposes into

\beq
\label{eq:rep}
H=H_0^{(r)} \oplus \cdots \oplus H_{F-1}^{(r)}
\eeq
where $H_i^{(r)}$ is the space on which $a_r^{\dag}$ has been applied
$i-$times. The fact that the state $|h>$ is annihilated by $a_r^{\dag}$
is legitimated by the property $\left(a_r\right)^F=0$.
\item
It is straightforward to check that the state
$$A_0 \cdots A_i |0>,$$
\noindent
is annihilated by $a_r$.
\end{enumerate}

Now we are ready to give the general ideas to built hermitian conjugate
operators. 
 We need first to define  number operators
$N^{(a)}_r$ and $N^{(b)}_s$ such that

\beqa
\label{eq:nop}
\begin{array}{ll}
[N_r^{(a)},a^{\dag}_s ] = \delta_{rs} a^{\dag}_s &
[N_r^{(a)},a_s ] = -\delta_{rs} a _s  \cr
[N_r^{(b)},b^{\dag}_s ] = \delta_{rs} b^{\dag}_s &
[N_r^{(b)},b_s ] = -\delta_{rs} b _s.
\end{array}
\eeqa 

In the case of infinite statistics as quons, those operators are complicated
polynomials which are expressed in terms of all the creation and annihilation 
operators \cite{numb}
and they contain terms of degree two, four, and so on.

Using those number operators, we define an alternative series of oscillators
$(a_r,a^{\dag}_r) \longrightarrow (\alpha_r, \alpha_r^{\dag})$ and
$(b_r,b^{\dag}_r) \longrightarrow (\beta_r, \beta_r^{\dag})$
(we just give the results for the $a$'s)
\beqa
\alpha^{\dag}_r&=& i q^{-1/2} a^{\dag}_r q^{-{N^{(a)}_r \over 4}} \\
\alpha_r &=&i q^{-1/2} q^{-{N^{(a)}_r \over  4}} a_r, \nonumber
\eeqa

\noindent
Then, using the properties of the algebra (and for instance its matrix 
realization),  we can prove

\beqa
\label{eq:q-quon}
q^{N^{(a)}_r} a_r &=& q a_r q^{N^{(a)}_r} \\
q^{N^{(a)}_r} a^{\dag}_r &=& q^{-1} a^{\dag}_r q^{N^{(a)}_r}. \nonumber
\eeqa
\noindent 
Next,
a direct calculation shows that the $\alpha$'s generate the $q-$oscillator
algebra introduced by Biedenharn and Macfarlane \cite{bm}

\beq
\label{eq:qosc}
%\left\{
%\begin{array}{ll}
\alpha_r \alpha_s^{\dag} - q^{1/2} \alpha_s^{\dag} \alpha_r = 
q^{-{N^{(a)}_r}/2} 
%\\
%\alpha_r \alpha_s^{\dag} - q^{-1/2} \alpha_s^{\dag} \alpha_r &= 
%(a_r a^{\dag}_r - a_r^{\dag} a_r)q^{-{N^{(a)}_r}/2}.
%\end{array}
%\right.
%\nonumber
\eeq

\noindent
It is then easy to build the Hilbert states from the $\alpha_r$,
when only one series of oscillators acts on the states
\beqa
\label{eq:qfock}
\alpha_r^{\dag} \left|k\right>&=& \sqrt{[k+1]} \left|k+1\right> \nonumber \\
\alpha_r \left|k\right>&=& \sqrt{[k]} \left|k-1\right> \\
{ N_r^{(a)}}  \left|k\right> &=& k \left|k\right>  \nonumber
%(a_r a_r^{\dag} - a_r^{\dag} a_r) \left|k\right> &=& k \left|k\right>,
%\nonumber
\eeqa

\noindent
with $[k] = {q^{k/2} - q^{-k/2} \over q^{1/2} -q^{-1/2}}$. 
Using explicitly the matrix realization of $\alpha_r, \alpha_r^{\dag}$
on this sub-space we see that the operator $\alpha_r$ and $\alpha_r^{\dag}$
are hermitian-conjugate. Therefore, we also get the conjugate
relation of (\ref{eq:qosc})
\beq
\label{eq:qosc2}
\alpha_r \alpha_s^{\dag} - q^{-1/2} \alpha_s^{\dag} \alpha_r = 
q^{{N^{(a)}_r}/2} 
\eeq

Consequently,  using the results established in the context of the
$q-$oscillators, we can see that the Hilbert space is definite positive
in this sub-space, and the representation  unitary.
The situation gets more involved when two different series of operators
act on the vacuum ($(\alpha_r^{\dag})^{k_r} (\alpha_s^{\dag})^{k_s}
|0>$ and so on). Therefore, we need certainly a more subtle transformation
than   (\ref{eq:q-quon}),  analogous to the complicated definition
of  number operators in the case of quons \cite{numb};
but this goes beyond the scope of this paper.\\
\no
There is a second difference connected to the non-commutativity 
of two operators of creation (or annihilation). Indeed, when
different series of operators are used we see that the number of
states increase with the degree of the monomial, and for instance
the two states $ \alpha_r^{\dag} \alpha_s^{\dag} |0> $ and 
$ \alpha_s^{\dag} \alpha_r^{\dag} |0> $ are different for $r \neq s$.

 All this structure, which turns out to be very complicated, can
certainly be interpreted using the quantum group limit of fractional
supersymmetry established in \cite{gq}, starting from a generic $q$, and
taking the limit $q \longrightarrow {\mathrm {~~primitive~~ root~~of~~unity}}$.

\mysection{Algebra and OPE}
\subsection{FSUSY algebra and OPE}

The Lagrangian (\ref{eq:faction}) is obviously invariant under conformal 
transformations. It 
turns out that $X(z)$ is of conformal weight $0$ and $\psi_{{a\over F}}(z)$ of 
conformal weight ${a \over F}$ as we will see. It is also invariant under the
FSUSY transformations generated by $Q$ (see (\ref{eq:fsusy})). In order to 
give the transformations, let us introduce 
$\epsilon$ the parameter of the FSUSY transformation. The $q-$mutation of
$\epsilon$ with $\psi_{{a \over F}}$ are identical as those of $\theta$ with
$\psi_{{a \over F}}$ as in Ref.\cite{d,prs}. This is a consequence of the FSUSY
transformation which corresponds to the translation $\theta \to \theta + 
\epsilon$ in the fractional superspace. The relation 
$\epsilon\theta=q^{-1}\theta\epsilon$ ensures that $D$ is 
a covariant derivative as it should be in order to build a FSUSY invariant 
action 
\cite{prs}. This can be proved in a straightforward  way using 
(\ref{eq:fsusyb}) which ensures that $\epsilon Q$ and $D$ commute. Then, the  
transformations of the superfield are 
$\delta_{\epsilon}\Phi=\epsilon Q\Phi$. They leave the Lagrangian invariant 
because the product of two superfields is a superfield and similarly for the covariant derivative. Using the decomposition of the field 
$\Phi$ this leads to
\beqa
\label{eq:trans}
\delta_{\epsilon}X&= &q^{{1\over 2}} \epsilon \psi_{{1\over F}}\nonumber\\
\delta_{\epsilon}\psi_{{a\over F}} &=& q^{{1\over 
2}}\{a+1\}\epsilon\psi_{{a+1\over 
F}},~~~~a\neq F-1 \\ 
\delta_{\epsilon}\psi_{{F-1\over F}}&=& (-1)^F q^{{1 \over 2}}
\epsilon\partial_{z}X,\nonumber
\eeqa
with $\{a\}={q^a-1\over q-1}$. 
Those transformation properties fit exactly (up to normalization
factors) with the ones introduced by Durand \cite{d}.\\
Stress that the coefficient of $\theta^{F-1}$ transforms as a total 
derivative. 
Therefore, with the rules of integration, the action is obviously FSUSY 
invariant. 
The generators of the conformal transformations (stress momentum tensor) and 
of the FSUSY transformations are
\beqa
\label{eq:T}
T(z) = -{1\over 2}:\partial_z X(z)\partial_z X(z):+{F\over (q-1)^F} 
\sum\limits_{a=1}^{F-1} &&\left[{F-a\over 2F}\left( 
(q^{-a}-1):\partial_z\psi_{{a\over 
F}}(z)\psi_{{F-a\over F}}(z)\right.\right.\nonumber\\
&&\left.\left.+(1-q^{a}):\psi_{{F-a\over F}}(z)\partial_z\psi_{{a\over 
F}}(z):\right)\right], 
\eeqa
\noindent 
and
\beq
\label{eq:G}
G(z) =q^{1/2}\left[:\partial_zX(z)\psi_{{1\over F}}(z):+
{F\over(q-1)^{F-1} }\sum\limits_{a=1}^{F-2}\{a+1\}\{-a\}{1\over 
1+\eta(a)}:\psi_{{F-a\over 
F}}(z)\psi_{{1+a\over F}}(z):\right].
\eeq 
\noindent
with $\eta(a)=q^{{F-1\over 2}}$ if $a={F-1\over 2}$ and $\eta(a)=1$ otherwise.
This complex normalization comes from the fact that we must distinguish in the 
OPE, the case $F-a=1+a$ from the other ones.
It appears that the stress-energy tensor, decomposes into ${F+1 \over 2}$ terms
which does not see each other because of the two-point Green functions 
(\ref{eq:green})
$$T(z) = \sum\limits_{a=0}^{{F-1 \over 2}} T_a(z),$$
with
\beqa
\label{eq:ta}
T_0(z) &=&  -{1\over 2}:\partial_z X(z)\partial X(z)  \\ 
T_a(z) &=& {F\over(q-1)^{F}}\left[{F-a\over 
2F}(q^{-a}-1):\partial_z\psi_{{a\over 
F}}(z)\psi_{{F-a\over F}}(z):+ {a\over 2F}(q^{a}-1):\partial_z\psi_{{F-a\over 
F}}(z)\psi_{{a\over F}}(z):\right.\nonumber \\ 
&+&\left.{F-a\over 2F}(1-q^{a}):\psi_{{F-a\over F}}(z)\partial_z\psi_{{a\over 
F}}(z):+ {a\over 2F}(1-q^{-a}):\psi_{{a\over F}}(z)\partial_z\psi_{{F-a\over 
F}}(z):\right]\nonumber\\
&& ~~~~~~~~~~~~~~~ a=1, \cdots,p={F-1 \over 2}. \nonumber 
\eeqa
Using the two-point correlation functions (\ref{eq:green}),  we have the 
following 
operator product expansion (OPE), encoding the different transformations 
\beqa
\label{eq:transf}
T(z) X(w) &=& {\partial_w X(w) \over z-w}+\cdots\nonumber\\
T(z) \psi_{{a\over F}}(w)&=& { {a\over F}\psi_{{a\over F}}(w) \over (z-w)^2}
+{\partial_w\psi_{{a\over F}}(w)\over z-w}+\cdots\nonumber\\
G(z) X(w) &=& q^{{1\over 2}}{\psi_{{1\over F}}(w)\over z-w}+ \cdots\\ 
G(z)\psi_{{F-1\over F}}(w)&=&(-1)^Fq^{{1\over 2}}{(1-q)\over 
F}^{F-1}{\partial_wX(w)\over 
z-w}+\cdots\nonumber\\
G(z)\psi_{{a\over F}}(w)&=&q^{{1\over 2}}\{a+1\}{\psi_{{a+1\over F}}\over 
z-w}+\cdots~.\nonumber
\eeqa
To compute these OPE, we have first used the Wick prescription detailed in 
subsection $4.1$, and then decomposed the fields in positive and negative 
modes.
As it was already mentioned, these OPE enable to fix in a consistent way the 
$q-$mutation
relations (\ref{eq:qab1},\ref{eq:qab2}) and therefore the relations 
(\ref{qchamp1},\ref{qchamp2}).
Moreover,
these transformations show explicitly that $X(z)$ is of conformal weight $0$,
and $\psi_{{a\over F}}$ of conformal weight ${a\over F}$. By comparing the 
OPE with the FSUSY transformations (\ref{eq:trans}), we conclude that
$G$ is the generator of the FSUSY transformations. FSUSY is then the natural 
extension of supersymmetry and connects ${1\over F}-$integer spin states. We 
have then to ensure that the algebra closes by computing the remaining OPE.

Mention that equations 
(\ref{eq:osc},\ref{eq:qab1},\ref{eq:qab2},\ref{eq:no}) which have been proven 
explicitly
for $F=3$, using
the oscillators $a$ and $b$, can be obtained along the same line for any $F$. 
Each
$\psi_{{ a \over F}},\psi_{{F- a \over F}}$ contribute to two series of 
oscillators
and   the relations equivalent 
(\ref{eq:osc},\ref{eq:qab1},\ref{eq:qab2},\ref{eq:no}) are  obtained with
the substitution $q \to q^a$. Then we get
\beqa
T(z) T(w) &=& {1\over 2} {c_F\over (z-w)^4} + {2 T(w)\over 
(z-w)^2}+{\partial_w 
T(w)\over z-w}+\dots\\
T(z) G(w) &=& {{F+1\over F}G(w)\over (z-w)^2}+ {\partial_wG(w)\over 
z-w}+\dots~.\nonumber 
\eeqa
\noindent
This shows that the conformal weights of $T$ and $G$ are $2$ and 
${F+1\over F}$ 
as it should be. The central charge is ($F>2$)
\beq
\label{eq:cf}
c_F= -12\sum\limits_{a=1}^{{F-1 \over 2}} \cos({2\pi  a\over F})
{a(F-a)\over F^2}. 
\eeq
\noindent
As  already established in \cite{prs}, the algebra does not close under
quadratic relations for $G$ with itself, because the underlying symmetries
involve $F-$power in the fractional superspace ($Q^F=\partial_z$).
This just tells us that there is,
in our case, no symmetry generator beyond $G$ and $T$ implying  
{\it non quadratic} closure relations as we will see.

The algebra we
are considering  closes upon $F-2$ intermediate would-be symmetry generators
$G_2(z),\cdots,G_{F-1}(z)$, with $G_{i}$ obtained from the OPE of $G$ upon
$G_{i-1}$. The reason why those operators do not generate a symmetry of the
action, has been analyzed with great details in \cite{prs}. Let us recall
briefly the main arguments. The symmetry of the Lagrangian is induced
by a basic symmetry in the superspace. The only operators (acting
in the superspace) which
satisfy the Leibniz rule, and thus generate a symmetry,
 are $\partial_z$ and $Q$. Consequently, only $T$
and $G$ generate a symmetry of the Lagrangian. This can be directly 
obtained, by observing that, the 
action of $G_{i \ne 1}$ on the fields do not leave the Lagrangian invariant. 
To summarize, the deep reason why the $G_{i \ne 1}$ are not symmetry operators
is a  reminiscence of the algebraic structure we are considering,
(which goes beyond (super-)Lie algebras). When $F$ is not a prime number, the 
situation is quite different and $G_i$ is a generator of symmetry if $i$ 
divides 
$F$ as will be seen further. \\

The conformal weight of these intermediate operators
is $1+{i \over F}$ and at the end of the process of closure, we have  the 
action
of $G$ on $G_{F-1}$. It leads to a tensor of conformal weight $2$, which can
be expressed as a sum of $T$ and possibly other terms expressed with the
various fields $X,\psi_{{a \over F}}$. This is a major difference with the 
fractional supervirasoro algebra where, due to the fractional power
appearing in the OPE, cuts are involved. 

As we have seen, 
using explicitly the modes expansions, one can built a representation of
the algebra starting from the vacuum previously defined. 
As in string theory, we
can consider different sets of sectors \cite{prs} depending on the boundary 
conditions of the fields. This constitutes an adapted generalization of
the Ramond-Neveu-Schwarz ones.
Therefore, the various quantum numbers of the vacuum can be calculated,
by regularizing the infinite sum by $\zeta$ function \cite{tye}
and an adapted GSO \cite{gso} projection, ensuring modular invariance,
as to be defined. The modes
of the two currents $G$ and $T$ allow to get a representation of the algebra
(\ref{eq:viral}) with a central extension (the last equation seems difficult
to be derived).
% and is probably related to the problem of associativity of the algebra). 
 This algebra can be compared with the 
fractional superconformal algebra
introduced in Ref. \cite{fv} which  is also generated, in addition to
the stress momentum tensor, by a current of conformal weight 
$(1 + {1 \over F})$.
These two extensions of the Virasoro algebra are different. 
The fractional superconformal algebra closes with rational 
power of $(z-w)$, leading to non-local algebras because cuts are involved.
The one we propose, closes only with {\it integer} power of $(z-w)$
but involves $F-th$ power instead of quadratic relations. 

Moreover, the central charges we get are  irrational numbers (except for
$F=2,3$). So the theory we have obtained  is 
no longer a rational conformal field theory (RCFT), but an irrational
conformal field theory (ICFT) (see \cite{icft} and references therein). 

\subsection{Associativity}
In a general CFT, there
are two strong requirements ensuring the consistency of the algebra, namely 
the 
closure relations and the associativity. The former just tells us that there 
is,
in our case, no symmetry generator beyond $G$ and $T$ implying  
{\it non quadratic} closure relations as we have seen before.
The latter is encoded through the
computation of  correlation functions which have to be invariant according to 
the way
we group the operators. This restricts considerably the possible CFT. 
For instance those constraints fixes the structure constants appearing when
considering the fractional supervirasoro algebras \cite{fsv} and is solved,
for instance, via the bootstrap equation coming from the four-point  
functions \cite{bpz}.  

Concretely, in order to set up the associativity, we have to compute
the four-point functions, when only the primary fields are involved.
For readability, we come back to the $F=3$ case.
Then, we define formally the OPE between the primary fields.
\beqa
\label{eq:opepf}
\partial X(z)\partial X(w)&=& {-1\over (z-w)^2}+\sum\limits_{n>0} 
C_{00}^{0,(n)} 
\partial^n X(w) (z-w)^{n-2}\nonumber\\
\partial X(z)\psi_{{a\over 3}}(w)&=&\sum\limits_{n\ge 0} C_{0a}^{a,(n)} 
\partial^n  \psi_{{a\over 3}}(w)(z-w)^{n-1},~~a=1,2\nonumber\\
\psi_{{1\over 3}}(z)\psi_{{1\over 3}}(w)&=&\sum\limits_{n\ge 0} 
C_{11}^{2,(n)}\partial^n\psi_{{2\over 3}}(w)(z-w)^n\\
\psi_{{2\over 3}}(z)\psi_{{2\over 3}}(w)&=&\sum\limits_{n\ge 0} 
C_{22}^{1,(n)}\partial^n\psi_{{1\over 3}}(w)(z-w)^{n-1}\nonumber\\
\psi_{{1\over 3}}(z)\psi_{{2\over 3}}(w)&=&{-q\over z-w}+\sum\limits_{n> 0} 
C_{12}^{0,(n)}\ \partial^n X(w)(z-w)^{n-1}\nonumber\\
\psi_{{2\over 3}}(z)\psi_{{1\over 3}}(w)&=&{q^2\over z-w}+\sum\limits_{n> 0} 
C_{21}^{0,(n)}\partial^n X(w)(z-w)^{n-1}\nonumber
\eeqa
At this point, the $C$ coefficients are unfixed but can be determined using 
constraints imposed by the associativity condition on correlation functions.\\

Therefore, from  the bootstrap equation, we have to calculate the four-point
function
$$<\phi_1(z_1) \phi_2(z_2)\phi_3(z_3)  \phi_4(z_4)>,$$ with $\phi_i$
an arbitrary primary field, in two different ways. The procedure is as follow.
Firstly, we do the
contraction of $\phi_1$ with $\phi_2$ and $\phi_3$ with $\phi_4$ and then
calculate the two-point functions. Secondly, we do the contraction
of $\phi_2$ with $\phi_3$ and $\phi_1$ with $\phi_4$ and determine the 
two-point 
functions. Equating the two ways of calculating the four-point function
gives some equations between the $C$'s. Technically, it is easier to use
the $SL(2,\CC)$ invariance of CFT to map $z_1 \to \infty$, $z_4 \to 0$
$z_2 \to 1$ and $z_3 \to x= {(z_1-z_2)(z_3-z_4) \over (z_2-z_4)(z_1-z_3)}$.\\
In our special case of FSUSY there is four types of four-point functions.\\
The first type concerns correlations functions involving only bosonic fields. 
In this case,
with the techniques described above,
the numbers $ C_{00}^{0,(n)}$ are those we find in standard conformal field 
theory.\\
The second type involves $4-$point functions like 
$$<\psi_{1 \over 3}(z_1) \psi_{1 \over 3}(z_2) \psi_{1 \over 3}(z_3)
\psi_{1 \over 3}(z_4)>$$ and similarly by changing $1\to 2$. The first way of 
doing the contractions gives zero.
When we use the second way, we get first
the three-point function 
$$<\psi_{1 \over 3}(z_1) 
\left(\sum\limits_{n\ge 0} C_{11}^{2,(n)}\partial^n\psi_{{2\over 3}}(w)(z-w)^n
\right)
\psi_{1 \over 3}(z_4)>.$$ But it seems impossible to make the contraction of
the first $\psi_1$ with the last one. However, if one calculate the three-point
function by doing one contraction we get also zero as it should be.\\
The third type of constraints are obtained from the calculation of correlation 
functions like 
$$<\psi_{1 \over 3}(z_1) \psi_{2 \over 3}(z_2) \psi_{1 \over 3}(z_3)
\psi_{2 \over 3}(z_4)>.$$ Using the bootstrap equations and (\ref{eq:opepf}), 
we thus obtain 
some constraints upon the $C_{12}^{0,(n)}$ and $C_{21}^{0,(n)}$.\\
And finally, the fourth type involves $4-$point functions mixing the fields
$\partial X$ and $\psi_{{a\over 3}}$. It imposes therefore constraints upon 
the  $C_{0a}^{a,(n)}$.

\no
The computation of all the $C$ coefficients is quite heavy and goes beyond the
 scope of this paper. Nevertheless, we have shown how it works in order to 
ensure the associativity of the algebra.

\mysection{Classification}
\subsection{Classification of FSUSY symmetries}

All the results established in the previous sections remain valid for 
arbitrary $F$, with some minor 
modifications. \\
If $F$ is not a prime number $F=fF'$, ($F'$ being a prime 
number)  we have $(\psi_{{f \over F}})^{F'}=0$ instead of $(\psi_{{f \over 
F}})^F=0$. In fact $\psi_{{f \over F}}$ is no longer a Grassmann variable of 
order $F$ but more precisely of order $F'$.

\hskip.3truecm If $F$ is a even number, the definition of $T_{{F \over 2}}$ in 
(\ref{eq:ta}) has  to be modified as 
$$ T_{{F \over 2}=F'} = -{2F \over (q-1)^F} \partial_z \psi_{{1\over 2}}(z) 
\psi_{{1\over 2}}(z), $$ 
and the central charge becomes 
$$c_F= 1 +  {1 \over 2}+ 2\sum\limits_{a=1}^{E({F-1 \over 2})} 
\cos({2\pi  a\over F}) \left\{ \left( {a \over F} \right)^2  +  
\left( {F-a \over F} \right)^2 -4 { a(F-a) \over F^2} \right\} , 
$$
where $E(~)$ means the integer part. In $T_{{F\over 2}}$, 
$\psi_{{1\over 2}}(z)$
is a usual fermionic field.  

\hskip.3truecm If $F$ is an odd number, the central charges (\ref{eq:cf}) and 
the stress momentum tensor  (\ref{eq:T}) remain unchanged.\\

\noindent

Some comments are in order here:  we can note that the central charge
is, in general, an irrational number but $F=2,3,4$. Among those families of 
theories, stress that for $F=4$ we do have the same central charge than for 
$F=2$. As a final remark we have, for $\psi_{{1 \over 2}}$ (when F even $\neq 
2$) a different normalization for $T_{{F \over 2}}$; this comes from the 
normalization in the action (\ref{eq:action}) and  in the Green function
(\ref{eq:green}).\\ 
The interesting point with those kinds of symmetries is that we are able to 
generalize the previous results  to any $F$ using its decomposition into prime
numbers. This exhibits, as we will see, substructures with smaller 
symmetries.
Let us consider the generic case when $F$ can be written as $F_1F_2$ with 
$F_1,F_2$ two  prime numbers not necessary different. A scalar multiplet of 
FSUSY  has the following irreducible decomposition in terms of $F_1$ multiplets
\beq
\Phi_F^{(0)} = \bigoplus\limits_{b=0}^{F_2-1}\Phi_{F_1}^{({b\over F})},
\eeq  
\noindent 
where $\Phi_{F_1}^{({b\over F})}$ is a $F_1$ multiplet of spin ${b\over F}$.\\
From this decomposition, it is obvious that one can get three different 
theories with the same fields. First,
using $\Phi_F^{(0)}$, we can built an  invariant FSUSY action. Furthermore,
 with the family of fields $\Phi_{F_1}^{({b\over F})}$ 
(or in the same way with 
$F_1\longleftrightarrow F_2$), a $F_1-$ ($F_2-$)SUSY can be also derived. 
However, the results are stronger because, by appropriate normalizations, the 
three Lagrangians so obtained are rigorously identical. This statement can 
be proved explicitly by a tedious calculation. We will just  give a
sketch of the proof and the exact normalizations will be omitted
for readability.
 The action (\ref{eq:action}) (which is 
also valid for arbitrary $F$ \cite{d}) can be reproduced using the fields 
$\Phi_{F_1}^{({b\over F})}$
\beqa
\Phi^{(0)}_{F_1} &\sim& X(z,\bar z) +  \sum \limits_{a=1}^{F_1-1}  
 \theta_1^a   \psi_{{aF_2\over F}}(z) 
\nonumber \\
\Phi^{({b \over F})}_{F_1} &\sim& \sum\limits_{a=0}^{F_1-1} 
 \theta_1^a \psi_{{aF_2+b\over F}}(z)\\
\Phi^{({F_2-b \over  F})}_{F_1}&\sim&\sum\limits_{a=0}^{F_1-1} 
 \theta_1^{F_1-a-1} \psi_{{F-aF_2-b
\over F}}(z)\nonumber \\ 
&&  ~~~~b=1,\cdots, E({F_2-1 \over 2}),  \nonumber
\eeqa
\noindent
where $\theta_1$ is a $F_1$ generalized Grassmann variable 
$(\theta_1^{F_1}=0)$
and $q_1=q^{F_2}$. The $q-$mutation of $\theta_{1}$ with the fields are given 
by
\beq 
\theta_1\psi_{{aF_2+b\over F}}=q^{-(aF_2+b)}\psi_{{aF_2+b\over F}}\theta_1.
\eeq
\noindent
The $aF_2-$components of the field $\Phi$ $q-$mute with the primitive root of
$F_1$ although the other ones $aF_2+b, b \neq 0$ with primitive root of $F$.
This is due to the non-trivial spin of the superfields 
$\Phi_{F_1}^{({b \over F})}$. Stress that $\theta_1$ is substituted to
$\theta$ in the $F_1$SUSY formulation.
From these relations and the normalization of the fields, we are now able to
write the $F_1$ SUSY invariant action 
\beq
{\cal L} \sim 
\int d\theta_1 \Bigg[
\partial_{\bar z}\Phi_{F_1}^{(0)}D_{F_1}\Phi_{F_1}^{(0)}+ 
\sum\limits_{b=1}^{E[{F_2-1\over 2}]} \left(
\partial_{\bar z}\Phi_{F_1}^{({b\over F})}\Phi_{F_1}^{({F_2-b\over F})}
+\partial_{\bar z}\Phi_{F_1}^{({F_2-b\over F})}\Phi_{F_1}^{({b\over 
F})}\right)\Bigg].
\eeq
\noindent
When $F_2=2$, the sum over $b$ contains just one term, namely $\partial_{\bar 
z} \Phi_2^{({1\over F})}\Phi_2^{({1\over F})}$. In principle heavy 
normalizations for the superfields have to be implemented in order to
reproduce the FSUSY action  (\ref{eq:action}).  \\
Consequently, if we have an action $F_1F_2$ supersymmetric, it is 
simultaneously $F_1$ and $F_2$ supersymmetric. And reciprocally, to get the 
converse, in addition to the scalar $F_1$ multiplet, we need ${1\over F}\dots
{F_2-1\over F}$ spin $F_1$ multiplets. The scalar will be coupled to itself 
via
the $F_1$ covariant derivative and the spin ${b\over F}$ with the spin 
${F_2-b\over F}$.  
Of course, as was already mentioned above, this result can be extended  by 
analogy for any F: we can conclude that in any case\\  
\vskip.1truecm
\centerline{
{ { if F' divides F, then a F--supersymmetric action is  
F'--supersymmetric.}}}
  
\vskip.6truecm
Along the same lines as for the FSUSY transformations, using the $F_1$SUSY
generator $Q_1= \partial_{\theta_1} + {(1-q_1) \over F_1}^{F_1-1}
\theta_1^{F_1-1} \partial_z$, 
we are able
to determine the $F_1$SUSY transformations : $ \delta_{\epsilon_1} 
\Phi_{F_1}^{({b\over F})} = \epsilon_1 Q_1 \Phi_{F_1}^{({b\over F})}$. 
Or in terms of the  components,

\beqa
&&\delta_{\epsilon_1} \psi_{{aF_2+b \over F}} \sim \epsilon_1 
\psi_{{(a+1)F_2+b \over F}},~~ a=0,\cdots, F_1-2  \\
&&\delta_{\epsilon_1} \psi_{{(F_1-1)F_2+b \over F}} \sim \epsilon_1 \partial_z 
\psi_{{b \over F}}, \nonumber
\eeqa

\noindent
where, for the sake of simplicity  we omit the normalizations for the 
superfield's transformations.
The transformations for $\Phi_{F_1}^{(0)}$ are similar to Eq.(\ref{eq:trans})
with $a \to aF_2$. Noticing that the spin of $\epsilon_1$ is $(-{1 \over F_1})$
one can easily check that both sides of the equation have the correct spin. 
In addition, as for  $\Phi_{F_1}^{(0)}$, the higher components of 
$\Phi_{F_1}^{({b\over F})}$, (with $b \neq 0$) transform as a total derivative.
This ensures that the previous  action, built up with the adapted superspace 
techniques, is automatically invariant under $F_1$SUSY transformations. To be 
as complete as possible, we give its generators
(omitting the normalizations)

\beqa
G_0(z) &\sim&  :\partial_zX(z)\psi_{{F_2\over F}}(z):  
+ \sum\limits_{a=1}^{F_1-2}:\psi_{{(F_1-a)F_2\over 
F}}(z)\psi_{{(1+a)F_2\over F}}(z):  \\
G_b(z)  &\sim& \sum\limits_{a=0}^{F_1-2} :\psi_{{F- aF_2 -b\over F}}(z)
\psi_{{(a+1)F_2 + b\over F}}(z): + :\psi_{{F_2 -b\over F}}(z) \partial_z 
\psi_{{b\over F}}(z) : ,~~~b=1,\cdots F_2-1. \nonumber
\eeqa

\noindent
Now, if we introduce  the $F_1-$multiplet of  spin ${b \over F}$  
$$(\psi_{{b \over F}}, \psi_{{b + F_2\over F}}, \cdots, \psi_{{b+aF_2  \over 
F}}, \cdots,  \psi_{{b+ (F_1-1)F_2  \over F}}),$$
and the $F_1-$one
of  spin ${F_2-b \over F}$  
$$(\psi_{{F_2-b \over F}}, \psi_{{ F_2-b+F_2  \over F}},  \cdots, 
\psi_{{F_2-b+ 
aF_2\over F}},\cdots,  \psi_{{F_2- b + (F_1-1)F_2 \over F}}),$$
we just see that
in the $G_b$ supercurrent, the fields appearing in the spin ${b \over F}$ 
multiplets couple the ones of the ${ F_2 - b \over F}$ multiplets. 
Using the Green functions (\ref{eq:green}), we get 
$$G_b(z) \psi_{{aF_2 +b} \over F}(w)\sim <\psi_{{F- aF_2 -b\over F}}(z)
\psi_{{aF_2 +b} \over F}(w)>\psi_{{{(1+a)F_2+b}\over F}}(z),~~a=0, \cdots, 
F-2,$$  
and
$$G_b(z) \psi_{{(F_1-1)F_2 +b} \over F}(w)\sim < \psi_{{F_2 -b\over F}}(z) 
\psi_{{(F_1-1)F_2 +b} \over F}(w)>
\partial_z\psi_{{b\over F}}(z). $$
We are then able to reproduce in a similar way the $F_1$SUSY 
transformations of $\Phi_{F_1}^{({b\over F})}$ with $G_b$. 
From this relation, we notice that we need simultaneously the fields 
$\Phi_{F_1}^{({b \over F})}$ and $\Phi_{F_1}^{({F_2-b \over F})}$   
for the supercurrent and the  action except for $b=0$.  
 
If one considers now the full action in two dimensions, with all the fields, 
the scalar $F$ superfield decomposes as
\beq
\Phi_F^{(0,0)}=\bigoplus\limits_{a,b=0}^{F_2-1}\Phi_{F_1}^{({a\over F},{b\over 
F})}.
\eeq
\noindent
We have four kinds of superfields: 

(i) $\Phi_{F_1}^{(0,0)}$ contains holomorphic, antiholomorphic and auxiliary
fields; 

(ii) $\Phi_{F_1}^{({a\over F},0)}$  holomorphic and auxiliary fields; 

(iii) $\Phi_{F_1}^{(0,{b\over F})}$ antiholomorphic and auxiliary fields; 

(iv)  $ \Phi_{F_1}^{({a\over F},{b\over F})}$ auxiliary fields. 

\subsection{Algebraic description}

To conclude those  series of inclusions, we can give an algebraic 
interpretation.
As we have mentioned previously, the underlying algebra of FSUSY is the one
generates by $\theta,\partial_\theta$. However, it is known that this algebra,
with the primitive root $q$, generates the $q-$deformed Heisenberg algebra
$H_q(q,\theta,\partial_\theta)$. If one considers the mapping ($F=F_1F_2$)

\beqa
f_2 :~ H_q(q,\theta,\partial_\theta) &&\to H_q(q,\theta,\partial_\theta) 
\nonumber \\
\theta &&\mapsto \theta_1=\theta^{F_2} \nonumber \\
\partial_\theta &&\mapsto \partial_{\theta_1} \\
q &&\mapsto q_1=q^{F_2}, \nonumber
\eeqa
\noindent
due to the fact that $f_2(\partial_\theta) = \partial_{\theta_1}$ and
$f_2(q) = q_1$ one  can check easily that $f_2$ is an homomorphism of algebra.
In this homomorphism, $\partial_{\theta_1}$ which is seen as a element of
$H_q(q,\theta,\partial_\theta)$, can be expressed as a polynomial of $\theta$
and $\partial_\theta$.
Then is we define the coset $H_q(q,\theta,\partial_\theta)/Ker(f_2)$ we 
get that this coset  is isomorphic to 
$H_{q_1}(q_1,\theta_1,\partial_{\theta_1})$.
Now, if we look at the $q-$mutation properties of the fields with $\theta_1$,
we have a third way to build the $F-$SUSY action. Using the $f_2$ 
isomorphism, we can identify $\theta_1$ with $\theta^{F_2}$. However, if we
proceed along those lines, the $q-$mutation relations (\ref{eq:qmut}) might
be incompatible with this identification. This appears when $F_1=F_2$, because
we cannot postulate simultaneously $\theta \psi_{{1\over F_1}} = q^{-F_1}
\psi_{{1\over F_1}} \theta$ and $\theta_1 \psi_{{1\over F_1}} = q_1^{-1} 
\psi_{{1\over F_1}} \theta$. Nevertheless, in such a situation, we can 
postulate only
the last $q-$mutation relation to  reproduce the action (\ref{eq:action}) 
with $F_1$ superfields by using appropriate normalizations.\\

\mysection{Conclusion}

We have constructed, in this paper, 
a conformal field theory using FSUSY. This conformal theory contains 
fractional spin states and is obtained 
analogously  as the superconformal algebra. This is achieved by the 
introduction
of an adapted Grassmann algebra. Those new variables encode the fractional spin
properties. After  quantization of the system, we obtain an Hilbert space 
where the quons  and the $q-$oscillators play a central role. It is worth
 noticing that supersymmetry is recovered when $F=2$. Quons and $q-$oscillators
 are just in this case the fermionic oscillators or the Clifford algebra. \\
The main feature of our algebra is that it closes through non-quadratic 
relations. Therefore, it cannot be seen as a Lie or super-Lie algebra. Due to
 the properties of these extended Grassmann variables, we have obtained a 
classification of the FSUSY algebra according to the decomposition of $F$ into 
its product of prime numbers. Next, according to this classification, we can wonder on the possible 
implications to have some sub-systems with smaller symmetries (when $F$ is not 
a prime number). Finally, we want to mention 
that the central charge obtained for $F=4$ is the same we get in 
supersymmetric theories.

This approach, as we have claimed, is different from the standard ones.
However, it should be interesting to have connections between this
way of doing and the standard affine Virasoro constructions (the  relations
of our model with the  Virasoro  master equation \cite{icft} has to be done).
Another open  (and related ?)  question concerns the relations of the basic 
fields  ($\psi_{{a \over F}}$) of conformal weight ${a \over F}$ with the 
parafermions introduced by 
Fateev and Zamolodchikov  \cite{fz}. Some clues have been given, and  as
we have already shown, it is possible to modify the algebraic structure in such
a way that  Green functions (\ref{eq:green}) with fractional
power of $(z-w)$ are involved \cite{prs}.

We have in our previous paper opened the possibility that such theories could
be the basic symmetry of the world-sheet of some string-inspired theory. We
have proved that the case $F=3$ leads to a rational critical dimension 
\cite{prs}. If we proceed along the same lines, it is easy to see that the
situation is less good for arbitrary $F \ne 2$.   When $F=4$ the anomaly
coming from SUSY and $4-$SUSY leads to a negative critical dimension,
and when $F \ge 5$ the critical dimension is irrationnal because of
the central charge which is irrational.

At this stage, only the cases $F=3,4$ should be
relevant for a relation with integrable systems, which has to
be established. For instance what kind of systems with $c=4/3-1=1/3$
should be described by $3-$SUSY ? 
However, $2D$ FSUSY  has the main advantage, even if relation with
integrable models is not obvious, that it
can be described using appropriate variables (the
generalized Grassmann variables).
This is in favor of a slight modification of our theory such that relations 
with  string or  integrable models are allowed. In this direction, 
two possible extensions of these results can be considered with a few changes.
The first one is to introduce interactions via an adapted superpotential. The 
second  extension can be performed by taking a superfield of conformal weight 
different from zero. This will clearly modify the values of the central 
charges. 
%Moreover, this has enabled to pass from a theory $F_1$-SUSY to a 
%theory $F_1\times F_2$-SUSY.

We can also  wonder on the possibility that each FSUSY extension of the 
Virasoro  algebra corresponds to a special point in some new series of 
integrable 
models {\it \`a la} Friedan-Qiu-Shenker.

We would also like to mention that the connection between FSUSY and the 
quantum groups has been undertaken recently \cite{gq}. In this paper, the 
authors  show
that, starting from arbitrary $q$, they get FSUSY in the limit where $q$ goes 
to
a primitive root of unity.

Finally, the ultimate dimension for the relevance of FSUSY is $D=1+2$. The $3D$
Poincar\'e algebra has been extended to a fractional supersymmetric extension.
Studying the representations of the corresponding algebra, relativistic anyons
are obtained \cite{rs}.\\

{\bf ACKNOWLEDGEMENTS}\\

\noindent
We would like to acknowledge P. Baseilhac, E. Dudas and V. Fateev for useful 
discussions  and  S.~-H.~Henry Tye for  correspondence and critical comments.

%\newpage

\end{document}